# Experimental verification of polar structures in ultrathin BaTiO$_3$ layers using resonant x-ray reflectivity

*Kook Tae Kim*[a‡], *Yeong Jae Shin*[b,c,d‡], *Sung-Jin Kang*[e], *Ryung Kim*[g], *Miyoung Kim*[e], *Tae Won Noh*[b,c], *Yongseong Choi*[f]\*, *Seo Hyoung Chang*[g]\*, *and Dong Ryeol Lee*[a]\*

[a]Department of Physics, Soongsil University, Seoul 06978, Republic of Korea

[b]Center for Correlated Electron Systems, Institute for Basic Science (IBS), Seoul 08826, Republic of Korea

[c]Department of Physics and Astronomy, Seoul National University, Seoul 08826, Republic of Korea

[d]Department of Materials Science Engineering, Yale University, New Haven 08826, United States

[e]Department of Materials Science and Engineering and Research Institute of Advanced Materials, Seoul National University, Seoul 08826, Republic of Korea

[f]Advanced Photon Source, Argonne National Laboratory, Argonne, Illinois 60439, USA

[g]Department of Physics, Chung-Ang University, Seoul 06974, Republic of Korea

**ABSTRACT**

Functional devices with ultrathin ferroelectric layers have been attracted as a promising candidate for next-generation memory and logic device applications. Using the ultrathin ferroelectric layers, particularly approaching two-dimensional limit, however, it is still challenging to control ferroelectric switching and to observe ferroelectricity by spectroscopic tools. In particular, conventional methods such as electrical measurements and piezoelectric response force microscopy are very limited due to leakage currents and the smallness of the ferroelectric signals. Here, we show that the ferroelectricity of ultrathin $SrRuO_3/BaTiO_3/SrRuO_3$ heterostructures grown on $SrTiO_3(100)$ substrates can be measured using resonant x-ray reflectivity (RXRR). This experimental technique can provide an element-specific electronic depth profile as well as increased sensitivity to Ti off-center displacements at the Ti K pre-edge. The depth-sensitivity of RXRR selectively detect the strong polarization dependence of the Ti pre-edge features of ultrathin $BaTiO_3$ layers while discriminating contribution of the $SrTiO_3$ substrate. This technique verified that the $BaTiO_3$ layer can be ferroelectric down to the lowest experimental limit of a critical thickness of 2.5 unit-cells. Our results can open a novel way to explore ultrathin ferroelectric-based nano-electronic devices.

INTRODUCTION

Ferroelectric materials have attracted tremendous interest due to their novel physical properties and potential nanoelectronic device applications. The electrically switchable spontaneous polarization of a ferroelectric layer enables electric-field control of its electrical transport, magnetic, mechanical, and optical properties.[1–4] Ferroelectric-based systems can be used as non-volatile memories,[5] ferroelectric field-effect transistors,[6] and integrated optical devices.[7–9] For the applications of nanoelectronics devices, it is quite important to understand the scaling of ferroelectric materials.[10–12] For instance, ultrathin ferroelectric tunnel junctions can exhibit tunneling electroresistance,[13–17] memristive switching,[18,19] and strong electromagnetic coupling, which have applications beyond complementary metal oxide semiconductor (CMOS) devices.[20,21] To explore these devices based on ultrathin ferroelectric materials, it is imperative to investigate ferroelectricity at ultrathin thickness limit.

The critical thickness of ferroelectric thin films has long been an important issue.[22–25] It is also prerequisite for the realization of beyond-CMOS devices with low power consumption. For instance, a theoretical study using the Landau-Ginzburg theory predicted the size effect of $BaTiO_3$ (BTO) at around 10-15 nm.[26] However, it has been experimentally observed that the critical thickness is thinner due to the development of the ferroelectric thin film growth method. Then, first-principles calculations proposed that a BTO film with top and bottom $SrRuO_3$ (SRO) electrodes can exhibit the ferroelectricity down to 3.5 unit cells (uc).[10,11] In this extremely thin regime, however, it is difficult to probe and verify the ferroelectric response by using the conventional electrical methos and polarization-field hysteresis loop measurements. As the ferroelectric layer thickness decreases to the nanoscale, leakage current significantly increases due to the unsolicited electron tunneling across the ferroelectric layer. Moreover, the piezoelectric displacement can be also quite small in the piezoelectric response force microscopy (PFM) measurements when an extremely small volume of ultrathin ferroelectric layer is used. The strong pinning effect of the ferroelectric domain walls with defects may interfere with PFM measurements. Therefore, despite numerous efforts, it remains challenging to detect weak ferroelectric signal experimentally.

Here, we report an experimental technique for investigating ferroelectric signals in ultrathin ferroelectric BTO films, namely polarization-dependent resonant x-ray reflectivity (RXRR). The BTO films sandwiched by metallic SRO electrodes were fabricated on $SrTiO_3$ (STO)(001) substrate using pulsed laser deposition. In the RXRR measurements, the pre-edge fine structure (PEFS) near Ti K absorption edge of the ultrathin BTO film was noticeably enhanced, evidencing the ferroelectricity of the BTO layer as thin as 2.5 uc. This is thinner than the previously reported experimental critical thickness for BTO ferroelectricity.[24] Bulk-sensitive x-ray absorption spectroscopy (XAS) measurements on these devices were not able to verify the ferroelectric distortion-related PEFS in Ti-K spectra[27–29] because of the strong contribution from the same Ti element in the STO substrate. RXRR can address an element specific electronic depth profiles.[30,31] This experimental technique opens the possibility of exploring ultrathin ferroelectricity beyond the conventional measurement limits.

RESULTS AND DISCUSSIONS

The ultrathin BTO films with metallic SRO top- and bottom-electrodes were grown by pulsed laser deposition on atomically smooth $TiO_2$-terminated STO (001) substrates. During the growth, the substrate temperature was maintained at 1000 K for both SRO and BTO growth. The ambient oxygen pressure was held at 100 mTorr and 5 mTorr for SRO and BTO, respectively. The thickness of each layer was monitored by using the specular spot intensity of reflection high-energy electron diffraction (RHEED) during the growth. **Figure 1**a,b shows the RHEED intensity profiles during the SRO/BTO/SRO heterostructure growths with 3 uc-thick SRO top- and bottom-electrode layers, respectively. For all bottom-SRO, BTO, and top-STO layers, the RHEED show clear two-dimensional patterns featured by sharp diffraction spots with streaky lines.[32] The clear oscillations observed in RHEED intensity profiles during BTO growth indicate that our BTO films are grown in two-dimensional layer-by-layer growth mode. The BTO growths are continued until four (Figure 1a) and three (Figure 1b) RHEED intensity oscillations are achieved, which yield BTO thickness ($t_{BTO}$) of 3.5 uc and 2.5 uc with BTO growth conditions used in this work.[33] For SRO growth, the RHEED intensity becomes saturated after the growth of two monolayers, indicating a growth mode transition from layer-by-layer growth to step-flow growth mode.[34]

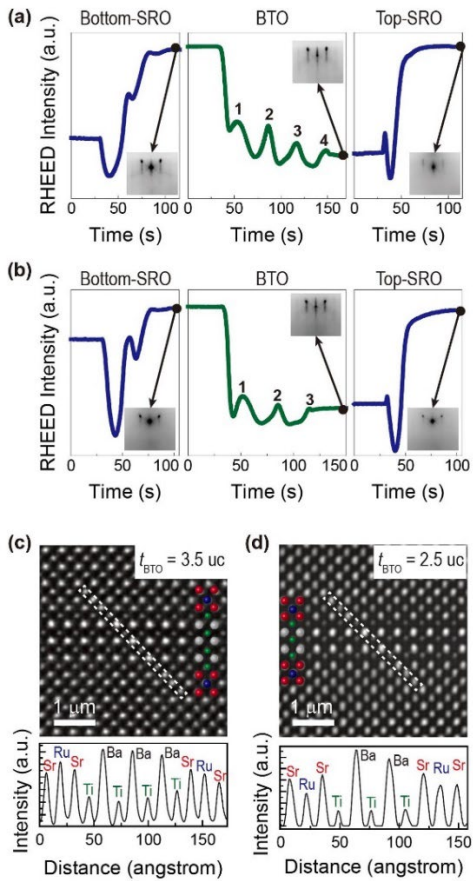

**Figure 1.** Time-dependent RHEED intensities for the growth of SRO/BTO/SRO heterostructures (a) with 3.5 uc-thick BTO layer and (b) with 2.5 uc-thick BTO layer. Insets are RHEED patterns of film surface taken along [100] azimuth. STEM-HAADF images of the SRO/BTO/SRO heterostructures with (c) 3.5 uc-thick BTO layer and (d) 2.5 uc-thick BTO layer along [100] zone axis. The following HAADF intensity profiles along the dashed boxes are also shown, respectively.

The atomic structures of SRO/BTO/SRO heterostructures are investigated using high-angle annular dark-field (HAADF)-scanning transmission electron microscopy (STEM). Figure 1c,d displays the HAADF images of 3.5 uc- and 2.5 uc-thick BTO films, respectively, with 20 nm-thick top- and bottom-SRO electrodes along the [100] zone axis. For both samples, atomically sharp interfaces with no significant defects were observed. The atomically sharp interface and the surface of our SRO/BTO/SRO heterostructures were further confirmed by atomic force microscopy (**Figure S1**). The HAADF intensity profile shows that the BTO layer consists of three (two) BaO layers and four (three) $TiO_2$ layers, confirming that $t_{BTO}$ = 3.5 uc (2.5 uc). Note that for both heterostructures with $t_{BTO}$ = 3.5 uc and $t_{BTO}$ = 2.5 uc, top and bottom interfaces are symmetrically terminated with SrO-$TiO_2$.

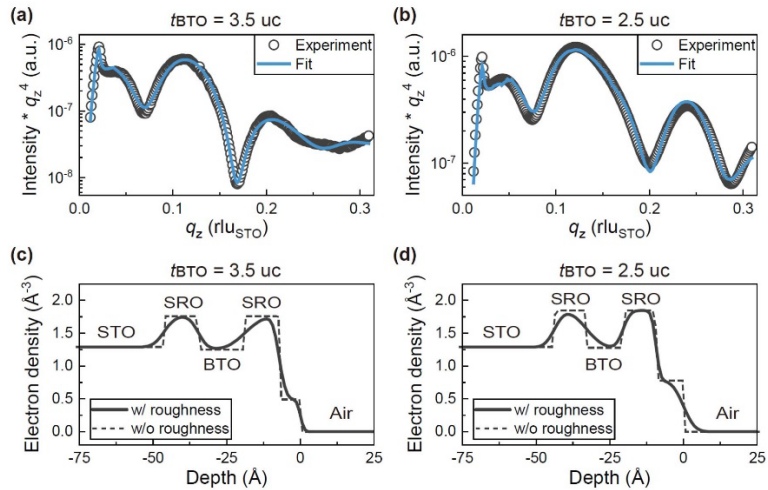

**Figure 2.** X-ray reflectivity of SRO/BTO/SRO heterostructures (a) with $t_{BTO}$ = 3.5 uc and (b) 2.5 uc measured at non-resonant energy, 4.96 keV (open circles). Electronic density profiles of SRO/BTO/SRO heterostructures (c) with $t_{BTO}$ = 3.5 uc and (d) 2.5 uc determined by the best fits to non-resonant x-ray reflectivity. The solid and dashed lines correspond to the electron density profiles obtained from the fitting with and without surface/interface roughness effects, respectively.

The structural properties of SRO/BTO/SRO heterostructures are further investigated using non-resonant x-ray reflectivity (see **Figure S2** for details of x-ray scattering geometry). Open circles in **Figure 2**a,b show the x-ray reflectivity measurements of SRO/BTO/SRO heterostructures for σ-polarized light as a function of the z component of the momentum transfer vector ($q_z$) measured at non-resonant energy (4.96 keV). As x-ray reflectivity is determined by the gradient of refractive index along the surface normal direction, and the total refractive index is proportional to the electron density, careful interpretation of the data provides electron density depth profile in sub-nanometer scale.[35] The x-ray reflectivity measurements were fitted using Parratt's recursive method.[36] The fitted curve (solid lines in Figure 2a,b) and experimental data show good agreement with each other. The solid lines in Figure 2c,d indicates the electron density profiles of SRO/BTO/SRO heterostructures obtained from the fit to x-ray reflectivity measurements.

Both for $t_{BTO}$ = 3.5 uc and $t_{BTO}$ = 2.5 uc, the estimated electron densities of each layer were well consistent with the values expected from the atomic form factors at measured energy (approximately 1.33, 1.76, and 1.30 electrons per unit-cell volume for BTO, SRO, and STO, respectively).[37] The thickness of each layer is determined from the electron density profiles based on the fit without surface/interface roughness effect considered (dashed lines). The estimated BTO thicknesses are about 14.1, and 11.1 Å for $t_{BTO}$ = 3.5 uc and $t_{BTO}$ = 2.5 uc, respectively, which are well-matched to values expected from layer thickness and bulk BTO c-axis lattice constant (4.038 Å). Note that our fitted electron density profiles indicate the existence of an additional low-density surface capping layer. This capping layer might originate from the low-density surface SRO layer[38] or ionic absorbates at the surface.[39]

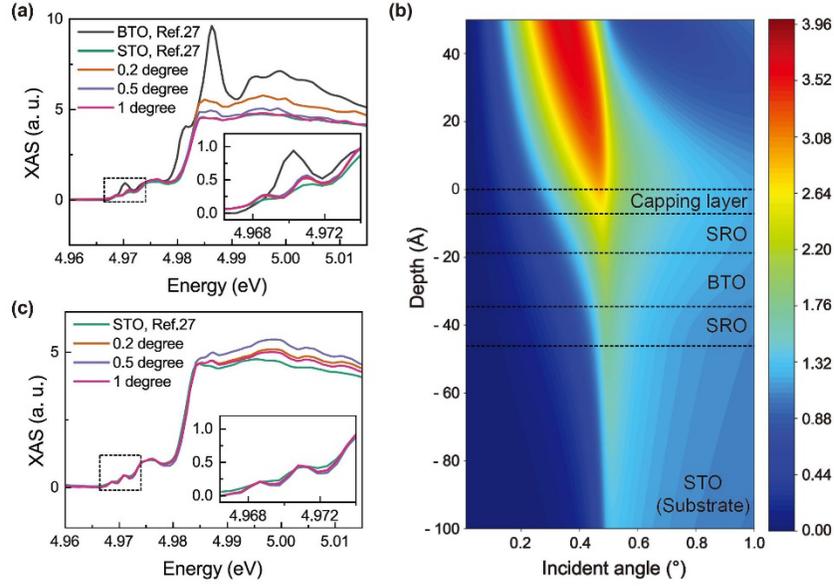

**Figure 3.** (a) Ti K-edge XAS of SRO/BTO/SRO heterostructure with $t_{BTO}$ = 3.5 uc on STO (001) for π-polarized light with incident angles of 0.2, 0.5, and 1 degree. The reference XAS for bulk STO and BTO crystals were also plotted.[27] (b) Electric field simulation result for SRO/BTO/SRO heterostructure with $t_{BTO}$ = 3.5 uc with respect to a depth from the surface and angle from the surface plane. The simulation was based on the electron density profile obtained in Figure 2(c). (c) Simulation results on Ti-K XAS of SRO/BTO/SRO heterostructure with incident angles of 0.2, 0.5, and 1 degree, based on the parameters obtained in (b) and reference XAS of bulk BTO and STO crystals. The reference XAS for bulk STO crystals are also plotted.[27]

The linear dichroism XAS has been one of powerful methods to quantitively measure the ferroelectricity of perovskite transition metal oxides. In particular, as the Ti ion is displaced from the centrosymmetric site within the TiO6 octahedra, the Ti pre-edge intensity is enhanced, proportional to the square of the displacement along the incident x-ray polarization.[Stern PRL 93, 037601] For perovskite titanates, it has been reported that the displacement between Ti with coplanar oxygen enhances the dipole-induced PEFS peak near Ti-K absorption edge (at about 4.97 keV) with the x-ray polarization parallel to the Ti displacement.[27–29] **Figure 3**a shows the Ti-K TFY-XAS of SRO/BTO/SRO heterostructure ($t_{BTO}$ = 3.5 uc) on STO (001) substrate for π-polarized light, with Ti K-edge absorption spectra of bulk STO and BTO crystals from Ref.27 overlaid for the comparison. The data was collected with incident angles ($\theta$) of 0.2, 0.5, and 1.5º to check grazing-incidence angle dependence of XAS for our SRO/BTO/SRO heterostructures.[40] For all incident angles, XAS of SRO/BTO/SRO heterostructure exhibits almost identical features with that of STO, while the ferroelectric BTO XAS shows a clear difference. The PEFS peak at around 4.97 keV presents a significant enhancement for bulk BTO due to the ferroelectric-related polar distortions of $TiO_6$ octahedral. On the other hand, no enhancement of PEFS peak is found for the paraelectric STO and SRO/BTO/SRO heterostructures. Considering that the ferroelectricity of the heterostructure with $t_{BTO}$ = 3.5 uc was confirmed by PFM measurements (see **Figure S3**a), the absence of enhancement in ferroelectric-related PEFS peak might be due to the strong contribution from Ti in paraelectric STO substrate, even with $\theta$ near the critical angle ($\theta$ ~ 0.5º, see Figure 3 and Section II in Supporting Information for more details).

To elucidate the absence of enhancement, total electric field amplitude inside the heterostructure was simulated using the parameters obtained from the fit to x-ray reflectivity in Figure 2c. Figure 3b represents the electric field simulation of SRO/BTO/SRO heterostructure as a function of depth from the surface and incident angles. Figure 3c shows the simulated Ti-K XAS of SRO/BTO/SRO heterostructures based on the electric field distribution in Figure 3b.[40] The calculated spectra in Figure 3c are well consistent with the experimentally measured XAS in Figure 3a for all measured $\theta$, exhibiting no enhancement of PEFS peak at 4.97 keV. The volume-averaged information provided by XAS is insensitive to the ultrathin BTO layer, due to the overwhelming contribution from the substrate Ti ions. Therefore, deeper understanding of ultrathin ferroelectric heterostructures requires individual layer- and depth-sensitive characterization method.

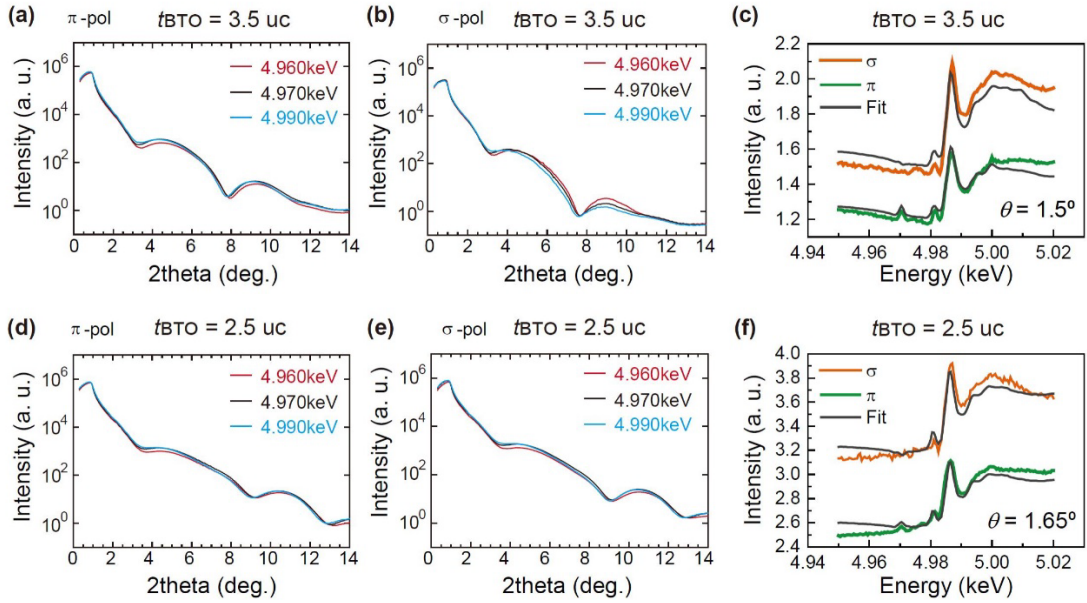

**Figure 4.** RXRR of SRO/BTO/SRO heterostructures with $t_{BTO}$ = 3.5 uc using (a) π-polarization (nearly perpendicular to the film plane), (b) σ-polarization (parallel to the film plane), and (c) energy-dependent x-ray reflectivity intensities at the incident angle of 1.5 degree. RXRR of SRO/BTO/SRO heterostructures with $t_{BTO}$ = 2.5 uc using (a) π-polarization, (b) σ-polarization, and (c) energy-dependent x-ray reflectivity intensities at the incident angle of 1.65 degree.

We performed RXRR measurements on the SRO/BTO/SRO heterostructures for selectively sensing the Ti K-edge absorption spectra of ultrathin BTO layers. Figure 4 represents the energy-dependent x-ray reflectivity intensities at near Ti K-edge with σ- (orange) and π-polarization (green). For both samples with $t_{BTO}$ = 3.5 uc and $t_{BTO}$ = 2.5 uc, all the data with π-polarized light exhibited an extra peak at around 4.97 keV, which does not appear in the XAS (Figure 3c) and the data with σ-polarized light. Even when tested near the grazing incident angle, conventional XAS measurements typically have a deeper penetration depth, e.g., 5 – 10 nm, than the overall thickness of our SRO/BTO/SRO layers, resulting in a dominent fluorescence signal from STO substrate. However, the RXRR measurement is sensitive to the interfacial contrast of electron, thus enhancing the sensitivity to the BTO layer. Furthermore, the layer contrast in the polarization dependence of the Ti-K PEFS, strong from the ultrathin BTO layer and absent from the STO substrate, allows isolating the Ti displacement that is linked to the BTO ferroelectricity.

RXRR measurements were quantitively analyzed for understanding the Ti-K PEFS and obtaining physical properties, such as dielectric functions, of SRO/BTO/SRO heterostructures. The refractive index with respect to measured energy $E$ and depth $z$ is given by

$$n(z, E) = 1 - \frac{\lambda^2 r_e}{2\pi} \sum_j \rho_{atom,j}(z)[f_j^{(1)}(E) + i f_j^{(2)}(E)] \tag{1}$$

where $\lambda$ is x-ray wavelength, $\rho_{atom,j}$ is the atomic density of layer $j$, $r_e$ is electron radius. $f_j^{(1)} + i f_j^{(2)}$ is the complex scattering factor of layer $j$, whose imaginary part is proportional to x-ray absorption.[41] To address the polarization-dependence of RXRR, we adopted the atomic scattering factor in tensor form. Due to its crystal symmetry, the scattering tensor of perovskite oxides is of the form[42,43]

$$\hat{f}_j(E) = \begin{pmatrix} f_{xx}(E) & 0 & 0 \\ 0 & f_{xx}(E) & 0 \\ 0 & 0 & f_{zz}(E) \end{pmatrix} \tag{2}$$

with $f_{\alpha\beta}(E) = f_{\alpha\beta}^{(1)}(E) + i f_{\alpha\beta}^{(2)}(E)$, where the $x$ and $z$ coordinate refer to the $a$ and $c$ crystallographic orientation. Note that $f_{xx}(E)$ and $f_{zz}(E)$ are related to σ- and π-polarized RXRR, respectively. Eqs. 1 and 2 imply that, by carefully interpreting polarization-dependent RXRR data, we achieved $f_{xx}^{(2)}$ and $f_{zz}^{(2)}$ spectra of each layer, and thus selective linear dichroism XAS of the ultrathin BTO layer.

The scattering factor of ultrathin BTO layer, $f_{BTO} = f_{BTO}^{(1)} + if_{BTO}^{(2)}$, was determined by simultaneous fitting of $q_z$-dependent non-resonant x-ray reflectivity curves (Figure 2) and energy-dependent RXRR data for all the $\theta$ angles (see Figure 4 and Figure S4 in Supporting Information). The non-resonant and resonant x-ray reflectivity curves are fitted using $f_{BTO}$ as a fitting parameter. The scattering factor of the STO layer was obtained by the reference STO Ti-K XAS (for the imaginary part) and its Kramers-Kronig transformation (for the real part). The SRO scattering factor was calculated by using reported atomic form factors of Sr, Ru, O atoms.[37]

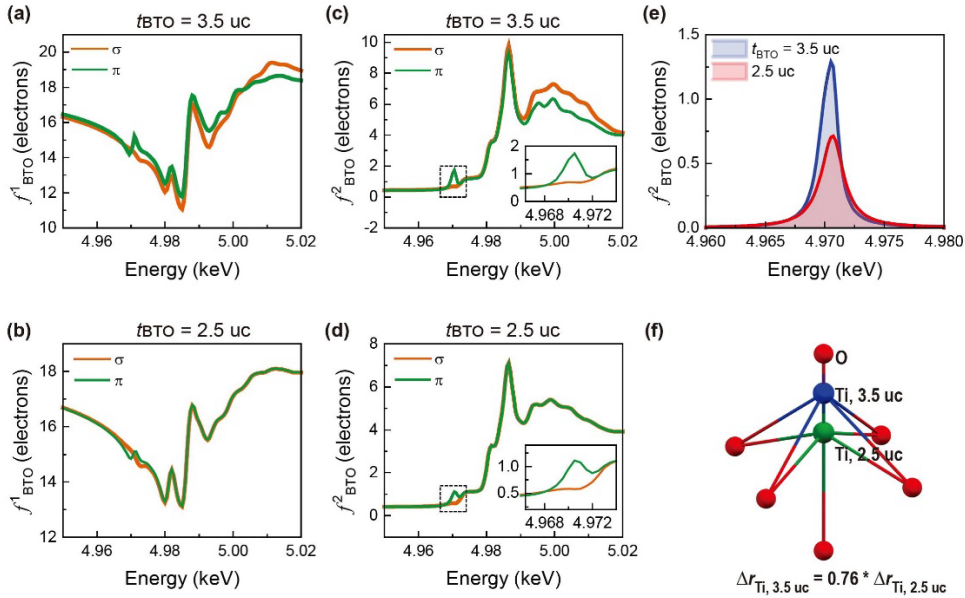

**Figure 5.** (a)[(b)] Real and (c)[(d)] imaginary part of atomic scattering factor of Ti in 3.5 uc (2.5 uc)-thick BTO layer. Ti scattering factor obtained from Ref. 36 is overlaid. (e) Intensities of Ti-K PEFS peaks for $t_{BTO}$ = 3.5 uc, and $t_{BTO}$ = 2.5 uc. (f) Schematic descriptions of ferroelectric distortions of TiO$_6$ octahedrons of 3.5 uc- and 2.5 uc-thick BTO layers.

The fitted RXRR curves (black curves in Figure 4) show good agreement with the experimental data. **Figure 5**a–d shows the scattering factors of ultrathin BTO layers. For the heterostructure with $t_{BTO}$ = 3.5 uc, the selective BTO Ti K-edge absorption spectrum (Figure 5c) exhibited a clear PEFS peak at around 4.97 keV for the π-polarized light, while absent for the σ-polarization. The strong polarization dependence reveals the Ti off-center displacement direction is along the depth direction and also indicates the ferroelectricity of the BTO layer. This was consistent with the PFM results on the sample with the same $t_{BTO}$ (Figure S3a). We tested the thinner BTO sample, which exhibit little ferroelectric hysteresis in the PFM measurement (see Figure S3b). Intriguingly, the ferroelectric-related PEFS peak was also detected in the heterostructure with $t_{BTO}$ = 2.5 uc (Figure 5d).

To get the further insight on the ferroelectric-related PEFS, we measured the PEFS peak area of both samples. The PEFS peak area is proportional to $\Delta r_{Ti}^2$, where $\Delta r_{Ti}$ is the relative displacement of Ti cations along the incident linear polarization direction from its centrosymmetric site within TiO6.. Figure 5e is the schematic, indicating the PEFS peak intensities of the samples. The measured PEFS peak area for $t_{BTO}$ = 2.5 uc was about 57 % of that for $t_{BTO}$ = 3.5 uc, indicating that the $\Delta r_{Ti}$ for $t_{BTO}$ = 2.5 uc was about 76% of that for $t_{BTO}$ = 3.5 uc (Figure 5f). The PEFS response is proportional to the square of the displacement along the x-ray polarization direction and thus insensitive to the sign of the displacement (i.e. two regions with positive and negative displacement will contribute equally) whereas the ferroelectric polarization of BTO via PFM is linearly proportional to net $\Delta r_{Ti}$.[44] The reduced ferroelectric polarization for $t_{BTO}$ = 2.5 uc might be explained by the increased depolarization field.[45] These results indicate that layer-selective absorption spectroscopy by RXRR can provide quantitative understandings on ultrathin ferroelectrics.

The inconsistency with the previous PFM study on 2.5 uc-thick BTO films could be explained by two possibilities: the reduction of piezoelectric displacement obtained from RXRR and ferroelectric polarization pinning. As total piezoelectric displacement is proportional both to the ferroelectric polarization and film thickness, the expected piezoelectric displacement for 2.5 uc-thick BTO can be significantly decreased about 50% of that for 3.5 uc-thick BTO rather than 24%. When the reduced piezoelectric displacement meets the detection limit of PFM, it is difficult to detect ferroelectric PFM hysteresis. Another possible explanation could be the ferroelectric polarization pinning. PFM measured the difference between oppositely aligned ferroelectric polarization, not absolute polarization value. Thus, PFM hysteresis would not be observed if the ferroelectric polarizations of 2.5 uc-thick BTO were not switchable because of interface dipoles or defects.[46,47]

CONCLUSIONS

In summary, we have investigated the polar structure of ultrathin ferroelectric heterostructures with 3.5 uc-thick and 2.5 uc-thick BTO layers using XAS and RXRR. The bulk-sensitive Ti-K XAS on ultrathin BTO heterostructures exhibited negligible enhancement of ferroelectric-related PEFS peak due to the dominant contribution from STO substrate. On the other hand, selectively achieved ultrathin BTO layer Ti-K spectra by RXRR shows clear enhancement in ferroelectric-related PEFS peak, providing quantitative information on ferroelectric polarization. Interestingly, using RXRR, we found that the ferroelectric $TiO_6$ distortion exists even for the 2.5 uc-thick BTO layer, which is thinner than the previously reported critical thickness for BTO ferroelectricity. These results suggest a novel way to investigate ferroelectricity in ultrathin limit.

EXPERIMENTAL SECTION

**Film growth.** SRO/BTO/SRO heterostructures were deposited on STO(001) substrates using pulsed laser deposition with a KrF excimer laser (248 nm, COMPex pro, Coherent). Before deposition, STO substrates are etched with a commercially available 10:1 buffered oxide etch for 30 s and annealed at 1100 ºC for 1 hour at the atmospheric pressure in order to achieve an atomically smooth $TiO_2$-terminated surface. During the deposition, laser fluence is set to 1.5 $J/cm^2$ for both BTO and SRO growth. The repetition rate of pulsed laser of 2 Hz and 3 Hz is used for BTO, and SRO growth, respectively. After the deposition, the substrate temperature is cooled down at a rate of 20 °C/min under an oxygen ambient of 100 mTorr.

**STEM measurements.** Atomic-scale imaging of ultrathin BTO heterostructures are obtained using STEM (JEM-ARM200F, JEOL) along the [001] zone axis. The background noise of obtained atomic images is removed using Average Background Subtraction Filter.[47]

**XAS and x-ray reflectivity measurements.** XAS and resonant and non-resonant x-ray reflectivity on ultrathin BTO heterostructures are measured at 4-ID-D beamline at Advanced Photon Source, Argonne National Laboratory. For XAS and energy-dependent RXRR, incident x-ray energy from 4.95 to 5.02 keV is scanned. The reflectivity measurements were carried out using an avalanche photodiode, and the XAS in fluorescence mode was measured using an energy dispersive detector collecting the Ti-$K_\alpha$ emission line.

ASSOCIATED CONTENT

**Supporting Information.** Atomic-force microscopy images of ultrathin BTO heterostructures, and experimental details of x-ray reflectivity analysis.

AUTHOR INFORMATION

**Corresponding Author**

*E-mail: ychoi@anl.gov

*E-mail: cshyoung@cau.ac.kr

*E-mail: drlee@ssu.ac.kr

**Author Contributions**

K.T.K. and Y.J.S. contributed equally. The manuscript was written through contributions of all authors. All authors have given approval to the final version of the manuscript.

**Notes**

The authors declare no competing financial interests.

ACKNOWLEDGEMENT

R. K. and S.H.C. were supported by Basic Science Research Programs through the National Research Foundation of Korea (NRF-2019K1A3A7A09033393, 2020R1A5A1016518, and 2020R1C1C1012424). Y.J.S. was supported by IBS-R009-D1 through the Research Center Program of the Institute for Basic Science in Korea. D.R.L and K.T.K were supported by National Research Foundation of Korea (NRF-2020R1A2C1009597 and NRF-2019K1A3A7A09033387). STEM analysis was supported by the National Center for Inter-University Research Facilities (NCIRF) in Seoul National University in Korea. This research used resources of the Advanced

REFERENCES

(1)  Dawber, M.; Rabe, K. M.; Scott, J. F. Physics of Thin-Film Ferroelectric Oxides. *Rev. Mod. Phys.* **2005**, *77* (4), 1083–1130.

(2)  Setter, N.; Damjanovic, D.; Eng, L.; Fox, G.; Gevorgian, S.; Hong, S.; Kingon, A.; Kohlstedt, H.; Park, N. Y.; Stephenson, G. B.; Stolitchnov, I.; Taganstev, A. K.; Taylor, D. V.; Yamada, T.; Streiffer, S. Ferroelectric Thin Films: Review of Materials, Properties, and Applications. *J. Appl. Phys.* **2006**, *100* (5).

(3)  Sando, D.; Yang, Y.; Paillard, C.; Dkhil, B.; Bellaiche, L.; Nagarajan, V. Epitaxial Ferroelectric Oxide Thin Films for Optical Applications. *Appl. Phys. Rev.* **2018**, *5* (4), 041108.

(4)  Chotorlishvili, L.; Jia, C.; Rata, D. A.; Brandt, L.; Woltersdorf, G.; Berakdar, J. Magnonic Magnetoelectric Coupling in Ferroelectric/Ferromagnetic Composites. *Phys. status solidi* **2020**, *257* (7), 1900750.

(5)  Scott, J. F. *Ferroelectric Memories*; Springer Series in Advanced Microelectronics; Springer Berlin Heidelberg: Berlin, Heidelberg, 2000; Vol. 3.

(6)  Kim, J. Y.; Choi, M.-J.; Jang, H. W. Ferroelectric Field Effect Transistors: Progress and Perspective. *APL Mater.* **2021**, *9* (2), 021102.

(7)  Abel, S.; Stöferle, T.; Marchiori, C.; Rossel, C.; Rossell, M. D.; Erni, R.; Caimi, D.; Sousa, M.; Chelnokov, A.; Offrein, B. J.; Fompeyrine, J. A Strong Electro-Optically Active Lead-Free Ferroelectric Integrated on Silicon. *Nat. Commun.* **2013**, *4* (1), 1671.

(8) Xiong, C.; Pernice, W. H. P.; Ngai, J. H.; Reiner, J. W.; Kumah, D.; Walker, F. J.; Ahn, C. H.; Tang, H. X. Active Silicon Integrated Nanophotonics: Ferroelectric BaTiO3 Devices. *Nano Lett.* **2014**, *14* (3), 1419–1425.

(9) Wessels, B. W. Ferroelectric Epitaxial Thin Films for Integrated Optics. *Annu. Rev. Mater. Res.* **2007**, *37* (1), 659–679.

(10) Gerra, G.; Tagantsev, A. K.; Setter, N.; Parlinski, K. Ionic Polarizability of Conductive Metal Oxides and Critical Thickness for Ferroelectricity in $BaTiO_3$. *Phys. Rev. Lett.* **2006**, *96* (10), 107603.

(11) Junquera, J.; Ghosez, P. Critical Thickness for Ferroelectricity in Perovskite Ultrathin Films. *Nature* **2003**, *422* (6931), 506–509.

(12) Kim, Y. S.; Kim, D. H.; Kim, J. D.; Chang, Y. J.; Noh, T. W.; Kong, J. H.; Char, K.; Park, Y. D.; Bu, S. D.; Yoon, J.-G.; Chung, J.-S. Critical Thickness of Ultrathin Ferroelectric $BaTiO_3$ Films. *Appl. Phys. Lett.* **2005**, *86* (10), 102907.

(13) Wen, Z.; Li, C.; Wu, D.; Li, A.; Ming, N. Ferroelectric-Field-Effect-Enhanced Electroresistance in Metal/Ferroelectric/Semiconductor Tunnel Junctions. *Nat. Mater.* **2013**, *12* (7), 617–621.

(14) Wang, L.; Cho, M. R.; Shin, Y. J.; Kim, J. R.; Das, S.; Yoon, J.-G.; Chung, J.-S.; Noh, T. W. Overcoming the Fundamental Barrier Thickness Limits of Ferroelectric Tunnel Junctions through $BaTiO_3$/$SrTiO_3$ Composite Barriers. *Nano Lett.* **2016**, *16* (6), 3911–3918.

(15) Garcia, V.; Fusil, S.; Bouzehouane, K.; Enouz-Vedrenne, S.; Mathur, N. D.; Barthélémy, A.; Bibes, M. Giant Tunnel Electroresistance for Non-Destructive Readout of Ferroelectric States. *Nature* **2009**, *460* (7251), 81–84.

(16) Garcia, V.; Bibes, M. Ferroelectric Tunnel Junctions for Information Storage and Processing. *Nat. Commun.* **2014**, *5* (1), 4289.

(17) Gruverman, A.; Wu, D.; Lu, H.; Wang, Y.; Jang, H. W.; Folkman, C. M.; Zhuravlev, M. Y.; Felker, D.; Rzchowski, M.; Eom, C.-B.; Tsymbal, E. Y. Tunneling Electroresistance Effect in Ferroelectric Tunnel Junctions at the Nanoscale. *Nano Lett.* **2009**, *9* (10), 3539–3543.

(18) Chanthbouala, A.; Garcia, V.; Cherifi, R. O.; Bouzehouane, K.; Fusil, S.; Moya, X.; Xavier, S.; Yamada, H.; Deranlot, C.; Mathur, N. D.; Bibes, M.; Barthélémy, A.; Grollier, J. A Ferroelectric Memristor. *Nat. Mater.* **2012**, *11* (10), 860–864.

(19) Kim, D. J.; Lu, H.; Ryu, S.; Bark, C. W.; Eom, C. B.; Tsymbal, E. Y.; Gruverman, A. Ferroelectric Tunnel Memristor. *Nano Lett.* **2012**, *12* (11), 5697–5702.

(20) Garcia, V.; Bibes, M.; Bocher, L.; Valencia, S.; Kronast, F.; Crassous, A.; Moya, X.; Enouz-Vedrenne, S.; Gloter, A.; Imhoff, D.; Deranlot, C.; Mathur, N. D.; Fusil, S.; Bouzehouane, K.; Barthelemy, A. Ferroelectric Control of Spin Polarization. *Science* **2010**, *327* (5969), 1106–1110.

(21) Yin, Y. W.; Raju, M.; Hu, W. J.; Burton, J. D.; Kim, Y.-M.; Borisevich, A. Y.; Pennycook, S. J.; Yang, S. M.; Noh, T. W.; Gruverman, A.; Li, X. G.; Zhang, Z. D.; Tsymbal, E. Y.; Li, Q. Multiferroic Tunnel Junctions and Ferroelectric Control of Magnetic State at Interface (Invited). *J. Appl. Phys.* **2015**, *117* (17), 172601.

(22) Qiao, H.; Wang, C.; Choi, W. S.; Park, M. H.; Kim, Y. Ultra-Thin Ferroelectrics. *Mater. Sci. Eng. R Reports* **2021**, *145*, 100622.

(23) Xie, L.; Li, L.; Heikes, C. A.; Zhang, Y.; Hong, Z.; Gao, P.; Nelson, C. T.; Xue, F.; Kioupakis, E.; Chen, L.; Schlom, D. G.; Wang, P.; Pan, X. Giant Ferroelectric Polarization in Ultrathin Ferroelectrics via Boundary-Condition Engineering. *Adv. Mater.* **2017**, *29* (30), 1701475.

(24) Shin, Y. J.; Kim, Y.; Kang, S.; Nahm, H.; Murugavel, P.; Kim, J. R.; Cho, M. R.; Wang, L.; Yang, S. M.; Yoon, J.; Chung, J.; Kim, M.; Zhou, H.; Chang, S. H.; Noh, T. W. Interface Control of Ferroelectricity in an SrRuO 3 /BaTiO 3 /SrRuO 3 Capacitor and Its Critical Thickness. *Adv. Mater.* **2017**, *29* (19), 1602795.

(25) Cheema, S. S.; Kwon, D.; Shanker, N.; dos Reis, R.; Hsu, S.-L.; Xiao, J.; Zhang, H.; Wagner, R.; Datar, A.; McCarter, M. R.; Serrao, C. R.; Yadav, A. K.; Karbasian, G.; Hsu, C.-H.; Tan, A. J.; Wang, L.-C.; Thakare, V.; Zhang, X.; Mehta, A.; Karapetrova, E.; Chopdekar, R. V.; Shafer, P.; Arenholz, E.; Hu, C.; Proksch, R.; Ramesh, R.; Ciston, J.; Salahuddin, S. Enhanced Ferroelectricity in Ultrathin Films Grown Directly on Silicon. *Nature* **2020**, *580* (7804), 478–482.

(26) Li, S.; Eastman, J.A.; Li, J.; Foster, C.M.; Newnham, R.E.; Cross, L.E. Size effects in nanostructured ferroelectrics. *Phys. Lett. A* **1996**, *212*, 341–346.

(27) Itié, J. P.; Couzinet, B.; Polian, A.; Flank, A. M.; Lagarde, P. Pressure-Induced Disappearance of the Local Rhombohedral Distortion in BaTiO$_3$. *Europhys. Lett.* **2006**, *74* (4), 706–711.

(28) Yoshii, K.; Yoneda, Y.; Jarrige, I.; Fukuda, T.; Nishihata, Y.; Suzuki, C.; Ito, Y.; Terashima, T.; Yoshikado, S.; Fukushima, S. Electronic Structure of $BaTiO_3$ Using Resonant X-Ray Emission Spectroscopy at the Ba-L3 and Ti-K Absorption Edges. *J. Phys. Chem. Solids* **2014**, *75* (3), 339–343.

(29) Vedrinskii, R. V.; Kraizman, V. L.; Novakovich, A. A.; Demekhin, P. V.; Urazhdin, S. V. Pre-Edge Fine Structure of the 3d Atom K x-Ray Absorption Spectra and Quantitative Atomic Structure Determinations for Ferroelectric Perovskite Structure Crystals. *J. Phys. Condens. Matter* **1998**, *10* (42), 9561–9580.

(30) Kemik, N.; Gu, M.; Yang, F.; Chang, C.-Y.; Song, Y.; Bibee, M.; Mehta, A.; Biegalski, M. D.; Christen, H. M.; Browning, N. D.; Takamura, Y. Resonant X-Ray Reflectivity Study of Perovskite Oxide Superlattices. *Appl. Phys. Lett.* **2011**, *99* (20), 201908.

(31) Hamann-Borrero, J. E.; Macke, S.; Gray, B.; Kareev, M.; Schierle, E.; Partzsch, S.; Zwiebler, M.; Treske, U.; Koitzsch, A.; Büchner, B.; Freeland, J. W.; Chakhalian, J.; Geck, J. Site-Selective Spectroscopy with Depth Resolution Using Resonant x-Ray Reflectometry. *Sci. Rep.* **2017**, *7* (1), 13792.

(32) Hasegawa, S. Reflection High-Energy Electron Diffraction. In *Characterization of Materials*; John Wiley & Sons, Inc.: Hoboken, NJ, USA, 2012.

(33) Shin, Y. J.; Wang, L.; Kim, Y.; Nahm, H.-H.; Lee, D.; Kim, J. R.; Yang, S. M.; Yoon, J.-G.; Chung, J.-S.; Kim, M.; Chang, S. H.; Noh, T. W. Oxygen Partial Pressure during Pulsed Laser Deposition: Deterministic Role on Thermodynamic Stability of Atomic Termination Sequence at $SrRuO_3$/$BaTiO_3$ Interface. *ACS Appl. Mater. Interfaces* **2017**, *9* (32), 27305–

27312.

(34) Rijnders, G.; Blank, D. H. A.; Choi, J.; Eom, C.-B. Enhanced Surface Diffusion through Termination Conversion during Epitaxial SrRuO3 Growth. *Appl. Phys. Lett.* **2004**, *84* (4), 505–507.

(35) Kim, D.-O.; Song, K. M.; Choi, Y.; Min, B.-C.; Kim, J.-S.; Choi, J. W.; Lee, D. R. Asymmetric Magnetic Proximity Effect in a Pd/Co/Pd Trilayer System. *Sci. Rep.* **2016**, *6* (1), 25391.

(36) Parratt, L. G. Surface Studies of Solids by Total Reflection of X-Rays. *Phys. Rev.* **1954**, *95* (2), 359–369.

(37) Henke, B. L.; Gullikson, E. M.; Davis, J. C. X-Ray Interactions: Photoabsorption, Scattering, Transmission, and Reflection at E = 50-30,000 EV, Z = 1-92. *At. Data Nucl. Data Tables* **1993**, *54* (2), 181–342.

(38) Shin, Y. J.; Lau, C.; Lee, S.; Walker, F. J.; Ahn, C. H. Surface-Induced Thickness Limit of Conducting La-Doped SrTiO 3 Thin Films. *Appl. Phys. Lett.* **2019**, *115* (16), 161601.

(39) Fong, D. D.; Kolpak, A. M.; Eastman, J. A.; Streiffer, S. K.; Fuoss, P. H.; Stephenson, G. B.; Thompson, C.; Kim, D. M.; Choi, K. J.; Eom, C. B.; Grinberg, I.; Rappe, A. M. Stabilization of Monodomain Polarization in Ultrathin PbTiO_{3} Films. *Phys. Rev. Lett.* **2006**, *96* (12), 127601.

(40) Souza-Neto, N. M.; Ramos, A. Y.; Tolentino, H. C. N.; Martins, A.; Santos, A. D. Depth-Dependent Local Structures in Thin Films Unraveled by Grazing-Incidence X-Ray

Absorption Spectroscopy. *J. Appl. Crystallogr.* **2009**, *42* (6), 1158–1164.

(41) *X-Ray and Neutron Reflectivity*; Daillant, J., Gibaud, A., Eds.; Lecture Notes in Physics; Springer Berlin Heidelberg: Berlin, Heidelberg, 2009; Vol. 770.

(42) LOVESEY, S.; BALCAR, E.; KNIGHT, K.; FERNANDEZRODRIGUEZ, J. Electronic Properties of Crystalline Materials Observed in X-Ray Diffraction. *Phys. Rep.* **2005**, *411* (4), 233–289.

(43) Dmitrienko, V. E.; Ishida, K.; Kirfel, A.; Ovchinnikova, E. N. Polarization Anisotropy of X-Ray Atomic Factors and `forbidden' Resonant Reflections. *Acta Crystallogr. Sect. A Found. Crystallogr.* **2005**, *61* (5), 481–493.

(44) *Physics of Ferroelectrics*; Topics in Applied Physics; Springer Berlin Heidelberg: Berlin, Heidelberg, 2007; Vol. 105.

(45) Mehta, R. R.; Silverman, B. D.; Jacobs, J. T. Depolarization Fields in Thin Ferroelectric Films. *J. Appl. Phys.* **1973**, *44* (8), 3379–3385.

(46) Liu, X.; Wang, Y.; Lukashev, P. V.; Burton, J. D.; Tsymbal, E. Y. Interface Dipole Effect on Thin Film Ferroelectric Stability: First-Principles and Phenomenological Modeling. *Phys. Rev. B* **2012**, *85* (12), 125407.

(47) Lee, D.; Jeon, B. C.; Baek, S. H.; Yang, S. M.; Shin, Y. J.; Kim, T. H.; Kim, Y. S.; Yoon, J.-G.; Eom, C. B.; Noh, T. W. Active Control of Ferroelectric Switching Using Defect-Dipole Engineering. *Adv. Mater.* **2012**, *24* (48).

(48) Kilaas, R. Optimal and Near-optimal Filters in High-resolution Electron Microscopy. *J.*

*Microsc.* **1998**, *190* (1-2), 45–51.

BRIEF

**Table of Contents graphic**

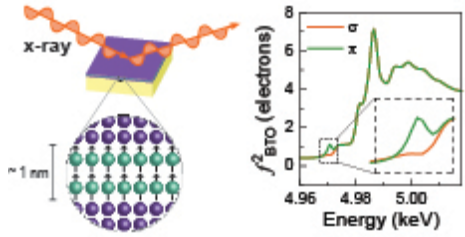